\documentclass{aastex}                     
\usepackage{spr-astr-addons}             
\usepackage[british]{babel}
\usepackage[latin1]{inputenc}
\usepackage{amsmath}

\begin{document}

\title{Dark matter as a condensate: Deduction of microscopic properties}

\shorttitle{Dark matter as a condensate}
\shortauthors{S. Guti\'errez, B. Carvente, A. Camacho}

\author{S. Guti\'errez}
\email{sergiogs@xanum.uam.mx}
\and
\author{B. Carvente}
\email{carvente@xanum.uam.mx}
\and
\author{A. Camacho}
\email{acq@xanum.uam.mx}
\affil{Departamento de F\'{\i}sica, Universidad Aut\'onoma Metropolitana--Iztapalapa\\
	Apartado Postal 55--534, C.P. 09340, M\'exico, D.F., M\'exico.}



\begin{abstract}
	 In the present work we model dark matter as a Bose--Einstein condensate
and the main goal is the deduction of the microscopic properties,
namely, mass, number of particles, and scattering length, related to
the particles comprised in the corresponding condensate. This task
is done with the introduction, in the corresponding model, of the effects of the
thermal cloud of the system. Three physical conditions are imposed,
i.e., mechanical equilibrium of the condensate, explanation of the
rotation curves of stars belonging to some galaxies, and, finally,
the deflection of light due to the presence of dark matter. These
three aforementioned expressions allow us to cast the features of
the particles in terms of detectable astrophysical variables.
Finally, the model is contrasted against the observational data of
fourteen galaxies and in this manner we obtain values for the involved
microscopic parameters of the condensate. The statistical errors are
five and twenty four percent for the scattering length and mass of the
dark matter particle, respectively.
\end{abstract}



\section{Introduction}

The current astrophysical observations allow us to conclude that
more than 27 percent of the content of the universe is comprised by
dark matter, dark energy near 68 percent, and baryons a 5 percent
\citep{Binney, Planck}. The conundrum appears when we note that
we do not see or understand the first two components. Here the
phrase dark matter can be understood as invisible, cold, and almost
pressureless matter distributed around the galaxies. In the context
of dark matter we may state that, in the gravitational realm, it
plays a dominant role, in other words, though it is a fundamental
element in the description of the Universe the knowledge that we
have of it is very scarce.

As mentioned before the corresponding evidence of its presence
emerges from astronomy \citep{Reid} and the rotation curves of 
galaxies provide one of the best examples of its existence.
Astronomical observations do attest that the rotational velocities
increase near the geometric center of the corresponding galaxy and
that these velocities remain almost constant. This last result
entails a mass profile for this kind of galaxies such that it does
not coincide with the content of baryonic mass.

There are very few facts that we know about it and they can be
mentioned very briefly. Indeed, if it were comprised by baryons the
cosmic microwave background would have a completely different
structure from the one observed, i.e., it cannot consist of baryons.
We must also discard light particles (not created by a phase
transition in the early Universe) \citep{Viel, Abazajian} and the
lower bound for the mass of the corresponding particle is
$m>2\mathrm{Kev}$. In addition, it has neither an electric charge
nor a magnetic dipole moment.

The lack of a deeper comprehension of the constituents of dark
matter has spurred a large number of models for its explanation.
Among the candidates we may find WIMPS (weakly--interacting massive
particles) \citep{Steigman}, axions \citep{Kolb}, etc. For a review on
this topic see \citep{Chavanis1} and references therein.

It is noteworthy to comment that among the aforementioned
possibilities we may found the option related to a Bose--Einstein
condensate (BEC) as a backbone description for dark matter
\citep{Baldeschi}. Here the physical meaning of a BEC is related to
the fact that at very low temperatures most of the particles of a
bosonic gas occupy the ground state of the corresponding
single--particle Hamiltonian operator \citep{Pethickbook}. In the
case in which only two--body interactions are taken into account the
BEC can be described resorting to the so--called Gross--Pitaevskii
equation, a non--linear Schr\"odinger differential equation in which
the non--linearity  has a cubic structure consequence of the
interaction among the particles of the gas \citep{Uedabook}.

The work in this direction includes the emergence of the seeds of
galaxies as a consequence of the presence of the BEC
\citep{Nishiyama}, a relativistic version of the Gross--Pitaevskii
equation \citep{Grifols}, BEC formed  by neutrinos due to the
assumption that they violate the Pauli Exclusion Principle
\citep{Dolgov}, etc. For a deeper look at the literature in this
context see \citep{Bohmer}.

It has to be emphatically underlined that in the deduction of the
Gross--Pitaevskii equation several assumptions are introduced,
namely, we neglect interactions between degrees of freedom related
to length scales smaller than the average interparticle spacing, and
all atoms are in the ground state of the corresponding Hamiltonian,
this is the so--called Hartree approximation \citep{Pethickbook},
only pairwise interactions are relevant. These conditions imply that
there are no particles in excited states, only the ground state is
populated. A few words concerning our model have to be said at this
point. Usually  in the analogy between dark matter and BEC
\citep{Chavanis1, Bohmer} the assumptions made involve that all
particles are in the ground state and that there is a scattering
length which is not necessarily null. The theory of ultra--cold
gasses shows us that if there is a non--vanishing scattering length,
then, inexorably, the number of particles populating excited states
cannot be zero \citep{Stringaribook1}, the so--called depletion term
appears. In the present work we take into account the thermal cloud
and analyze its consequences.

The mathematical model is defined by Gross--Pitaevskii--Poisson
system (see below) \citep{Chavanis1, Bohmer} the one, in the realm of
analytical results, is usually accompanied by the Gaussian Ansatz,
i.e., an approximate expression for the density profile is given in
the form of a Gaussian function.

In the present work we will provide an argument which explains the
physics behind this assumption as a consequence of the fact that in
the roughest approximation the self--gravitational interaction of
the halo can be understood as the case of a BEC immersed in an
isotropic harmonic oscillator trap. It will be shown that if we
follow this line of argumentation we may recover, immediately, the
radius of a BEC (for the case in which no scattering length is
present) without imposing any additional requirement. This
interpretation will endow the system with an effective trapping
potential and, in consequence, we will deduce the energy of the gas
assuming that only the ground and the first excited states are
populated.

The extant work in this context usually considers the consequences
of a BEC upon the rotation curves of galaxies, the appearance of
gravitational instabilities, etc. In other words, the microscopic
properties of the BEC are assumed and starting from this primordial
assumption the corresponding consequences upon diverse physical
scenarios are obtained. It has to be underlined the fact that (as
far as the authors know) there is a void in the literature in this
topic. Indeed, there are no works in which the microscopic
properties of a BEC, depicting dark matter, are deduced from the
current observational data. This is a task that has to be done since
amidst the scarce information concerning dark matter at our
disposition we may find a lower bound for the mass of the involved
dark matter particle. This last comment defines the main goal of the
present work, namely, the deduction of the microscopic properties
(mass, scattering length and number of particles) in a BEC in the
context of dark matter. Here we do not assume anything about the
mass or scattering length of the dark matter particles, but deduce
them from the current observational data.

Since three physical parameters are to be obtained we need, at
least, three independent expressions relating the microscopic
properties to variables obtained by observational data. In order to
have these required expressions we impose three physical conditions
upon our BEC: (i) mechanical equilibrium of the condensate; (ii)
concordance of the model with the rotation curves of stars belonging
to some galaxies, and, finally; the deflection of light due to the
presence of dark matter. Let us now explain, briefly, how we will
proceed. Firstly, the energy of the gas will be calculated and with
it the condition of mechanical equilibrium deduced, a fact leading
us to a relationship defining the radius of the BEC as a function of
the mass and scattering length of the dark matter particles and,
also as a function of the number of particles in the condensate.
Secondly, we relate the tangential velocity of galaxies with the
matter content of our halo. Finally, we consider a light beam moving
toward the halo and the ensuing deflection due to the mass of the
condensate is obtained. Clearly, the model contains three unknown
features related to the BEC, i.e., mass and scattering length of the
dark matter particles plus the number of particles in the gas and
our work includes three features which can be detected by
astrophysical and astronomical means and depending upon our BEC
characteristics.

We apply our model to the case of some galaxies and find that the
mass and scattering length of the dark matter particle are, $m\sim
10^{1}\mathrm{eV}/c^{2}$, $a\sim 10^{-6}\mathrm{m}$, respectively. In
addition, we will calculate the number of particles in the BEC for
the chosen cases. Since the elastic cross section (at low energies)
is a quadratic function of the scattering length $\sigma = 4\pi a^2$
\citep{Cohenbook2} then our results entail an elastic cross section
equal to $\sigma\sim 10^{-11}\mathrm{m}^2$. Finally, it will be
shown that the ensuing statistical errors in the deduction of the
values for $a$ and $m$ are five and twenty four percent, respectively.

Summing up, the main result from our work states that the current
observational data allow us to deduce the microscopic properties of
a dark matter halo modeled as a BEC.

\section{Dark Matter and BEC}

\subsection{Gravitational Interaction and Trapping Potential}

As mentioned before we consider $N$ dark matter particles each one
of them with mass $m$ and scattering length $a$ and forming a
spherical body with radius $R$. This symmetry is a consequence of
the assumption that all kind of external interactions upon the halo
are neglected in the present work and, therefore, no privileged
direction appears and hence spherical symmetry is the only possible
option for the geometry of the halo.

The usual work in this context considers the
Gross--Pitaevskii--Poisson equation, nevertheless the analytical
case requires (due to the fact that there are no exact solutions to
this situation) an initial assumption for the mass density of the
dark matter and the choice is usually a Gaussian function
\citep{Chavanis1}. Concerning this aforementioned condition we
proceed to provide a physical argument that explains this functional
structure as a consequence of the self--gravitational interaction of
the halo.

Consider a spherical body of mass $M$ and radius $R$ the one has a
small cavity along the diameter coincident with the $z$--axis. In
addition, it has a density function with spherical symmetry, i.e.,
it is only a function of the radial variable. Notice that in the
present situation the density has a non--compact support. We now
proceed to prove that it has to be depicted by a function such that
its Taylor expansion about the center of the body renders a series,
not a polynomial. Indeed, we proceed by contradiction, therefore,
let us assume that

\begin{eqnarray}
	\rho(r)= \rho(0)+\sum_{n=1}^{\infty}\rho^{(n)}(0)r^n/n!.\label{equa01}
\end{eqnarray}

Such that $\exists~s\in \mathbb{N}\backepsilon \rho^{(n)}(0)=0,
\forall~n\geq s$. This condition implies that around the center of
the body the density is a polynomial and not a series. Therefore

\begin{eqnarray}
	r\rightarrow\infty\Rightarrow\lim\rho(r)\rightarrow\infty .\label{equa02}
\end{eqnarray}

Clearly, this is a non--physical situation and it implies that the
density cannot be depicted by a polynomial. In other words,
$\forall~l\in\mathbb{N}, \exists~s\in\mathbb{N}$ such that
$\rho^{(s)}(0)\not=0, s\geq l$, i.e., it is a series. Physically the
meaning of this result is that far away from the center of the body
the density vanishes, a fact impossible to achieve with a
polynomial.

We now proceed to find an approximate expression for the motion of
our particle moving along the diameter coincident with the
$z$--axis. In order to do this let us consider one condition upon
$\rho(r)$, namely, its global maximum lies at $r=0$. We may fathom
this condition from a physical perspective  noting that, from an
intuitive point of view, gravity entails that we expect the maximum
of the density to be at the geometrical center of the body.

A particle of mass $m$ moves along this cavity and we determine the
classical motion equation for $m$. The coordinate system has its
origin at the geometrical center of the body, and we resort to
spherical coordinates. Then the motion equation reads

\begin{eqnarray}
	m\frac{d^2r}{dt^2}=-G\frac{mM(r)}{r^2}.\label{equa03}
\end{eqnarray}

Where $M(r)$ denotes the mass inside the sphere of radius $r$. Due
to our assumptions

\begin{eqnarray}
	M(r)=\frac{4\pi}{3}\rho(0)r^3\left[1+3\sum_{n=2}^{\infty}\frac{\rho^{(n)}(0)}{\rho(0)n!(n+3)}r^n\right].\label{equa04}
\end{eqnarray}

The equation of motion reads

\begin{eqnarray}
	\frac{d^2r}{dt^2}+\frac{4\pi}{3}G\rho(0)r\left[1+3\sum_{n=2}^{\infty}\frac{\rho^{(n)}(0)}{\rho(0)n!(n+3)}r^n\right]=0.\label{equa05}
\end{eqnarray}

Since the series in our last expression exists then we may define

\begin{eqnarray}
	f(r)\equiv\sum_{n=2}^{\infty}\frac{\rho^{(n)}(0)}{\rho(0)n!(n+3)}r^n.\label{equa06}
\end{eqnarray}

As a consequence of these arguments we have that $\forall~\delta >0,
\exists~l\in\mathbb{N}$ such that

\begin{eqnarray}
	\left| f(r)-\sum_{n=2}^{s}\frac{\rho^{(n)}(0)}{\rho(0)n!(n+3)}r^n\right| <\delta,~\forall~s\geq l.\label{equa07}
\end{eqnarray}

This last condition implies

\begin{eqnarray}
	\left|\sum_{n=s+1}^{\infty}\frac{\rho^{(n)}(0)}{\rho(0)n!(n+3)}r^n\right| <\delta,~\forall~s\geq l.\label{equa08}
\end{eqnarray}

If $\delta_1=10^{-1}$, then $\exists ~l_1\in\mathbb{N}$ such that

\begin{eqnarray}
	\left|\sum_{n=s+1}^{\infty}\frac{\rho^{(n)}(0)}{\rho(0)n!(n+3)}r^n\right| <10^{-1},~\forall~s\geq l_1.\label{equa09}
\end{eqnarray}

This last expression entails an inequality for our equation of
motion, indeed

\begin{eqnarray}
	-\frac{d^2r}{dt^2}\leq\frac{4\pi}{3}G\rho(0)r\left[\frac{13}{10}+3\sum_{n=2}^{l_1}\frac{\rho^{(n)}(0)}{\rho(0)n!(n+3)}r^n\right].\label{equa10}
\end{eqnarray}

At this point we consider that our density function has the
following characteristic

\begin{eqnarray}
	\left|\frac{\rho^{(n)}(0)}{\rho(0)}\right|\sim\frac{1}{R^n}.\label{equa11}
\end{eqnarray}

In this last expression $R$ is the characteristic radius of the
region comprising most (at least 87 percent) of the particles. A
word of warning is, at this point, noteworthy. Indeed, it can be
easily checked that this feature is shared by several non--compact
(integrable over the whole space) functions, for instance, Gaussian,
Lorentzian functions and also all the even eigenfunctions of a
harmonic oscillator.

Since the motion of our test particle takes place within the region
defined by $r\in[0,R]$ then the ensuing equation of motion has the
following approximate expression

\begin{eqnarray}
	\frac{d^2r}{dt^2}+\frac{4\pi}{3}G\rho(0)r\left[\frac{13}{10}+3\sum_{n=2}^{l_1}\left(\frac{r}{R}\right)^n\right]=0.\label{equa12}
\end{eqnarray}

Notice that the term with the factor $13/10$ defines a three
dimensional harmonic oscillator and the additional ones may be
understood as perturbations to it, at least in those cases in which
$r<R$. The corresponding frequency is given by

\begin{eqnarray}
	\omega_{(0)}=\sqrt{\frac{52\pi G\rho(0)}{30}}.\label{equa13}
\end{eqnarray}

This expression tells us that in a very rough approximation the
motion of $m$ is related to an isotropic harmonic oscillator whose
frequency depends upon the central density. With this classical
result we now quantize our system and consider the gravitational
effects of $N-1$ particles of mass $m$ ($M=(N-1)m$) upon our
particle of, also, mass $m$. The corresponding time--independent
Schr\"odinger equation is then

\begin{eqnarray}
	E\psi=-\frac{\hbar^2}{2m}\nabla^2\psi +\frac{m\omega^2_{(0)}}{2}r^2\psi .\label{equa14}
\end{eqnarray}

Clearly, the ground state of this last equation is a Gaussian
function \citep{Cohenbook1}. Since the Hartree approximation is
introduced then all the particles are described by a Gaussian
wavefunction and therefore the mass density shares this feature. In
other words, these previous arguments show us that this ubiquitous
Gaussian Ansatz can be understood as an approximation in which
gravity defines a three--dimensional harmonic oscillator. There is
an additional lesson to be elicited from these arguments, namely,
the self--gravitational effects of the halo can be, in a first and
rough version, be considered as defining an external trapping
isotropic harmonic oscillator.

Furthermore we may obtain the radius (radius means the distance from
the center of the halo such that the derivative of the rotational
speed of galaxies vanishes). It is noteworthy to mention that within
this region lie most of the particles, i.e., $87$ percent of them.
Indeed, notice that the ground state wave function related to
(\ref{equa14}) is a Gaussian function and it provides us the mass
density

\begin{eqnarray}
	\rho(r)=\frac{mN}{\sqrt{\pi^3}l^3}\exp(-r^2/l^2),\label{equa15}
\end{eqnarray}

\begin{eqnarray}
	l=\sqrt{\frac{\hbar}{m\omega_{(0)}}},\label{equa16}
\end{eqnarray}

\begin{eqnarray}
	\rho(0)=(0.16)\frac{mN}{l^3}.\label{equa17}
\end{eqnarray}

Joining this definition and the relation for $l$ and the frequency
entails

\begin{eqnarray}
	\omega_{(0)}=\sqrt{\frac{0.9GmN}{l^3}}.\label{equa18}
\end{eqnarray}

\subsection{Mathematical model}

We now take into account the case in which a non--vanishing
scattering length is present in the system as an essential element.
As mentioned before, the self--gravitational interaction of the dark
matter halo will be considered as an isotropic harmonic oscillator
and therefore our system is reduced, in its mathematical analysis,
to the case of a terrestrial BEC \citep{Pethickbook}.

According to our interpretation of the self--gravita\-tional
interaction of the halo as an external isotropic three--dimensional
harmonic oscillator we have that the corresponding many
body--Hamiltonian, for the situation of a dilute gas, is given by

\begin{eqnarray}
	\hat{H}=\sum_{i=1}^N\left[\frac{p_{(i)}^2}{2m}+\frac{m\omega^2_{(0)}}{2}r_{(i)}^2\right]\nonumber\\
	+U_{(0)}\sum_{i<j}\delta\left(\vec{r}_{(i)}-\vec{r}_{(j)}\right).\label{equa19}
\end{eqnarray}

In this last expression the term $U_{(0)}=4\pi\hbar^2a/m$ contains
the information, under the assumption of very low energies and
dilute gas, of the interaction between two particles. In other
words, the present model assumes, due to the fact that we have a
very dilute gas, that only pairwise interactions are relevant for
the dynamics of the system \citep{Pethickbook}. Here $a$ denotes the
scattering length of our particles. In this so--called mean--field
treatment all short--wavelength degrees of freedom have been
integrated out \citep{Uedabook}.

The time--independent equation in the context of the Hartree
approximation is the so--called Gross Pitaevskii equation and reads

\begin{eqnarray}
	\mu\psi(\vec{r})= \left[-\frac{\hbar^2}{2m}\nabla^2 +\frac{m\omega^2_{(0)}}{2}r^2 +U_{(0)}\vert\psi(\vec{r})\vert^2\right]\psi(\vec{r}),\label{equa20}
\end{eqnarray}

\begin{eqnarray}
	N=\int\left|\psi(\vec{r})\right|^2d\vec{r}.\label{equa21}
\end{eqnarray}

In the equation of motion $\mu$ denotes the chemical potential.
Since $\rho(\vec{r})=m\vert\psi(\vec{r})\vert^2$ we may cast
(\ref{equa20}) in the following form

\begin{eqnarray}
	\mu\psi(\vec{r})= \left[-\frac{\hbar^2}{2m}\nabla^2 +\frac{m\omega^2_{(0)}}{2}r^2 +\frac{U_{(0)}}{m}\rho(\vec{r})\right]\psi(\vec{r}).\label{equa22}
\end{eqnarray}

The lack of analytical solutions in this realm \citep{Liebbook}
obliges us to introduce some approximation and at this point it is
noteworthy to comment the assumptions that can be found in the
literature and compare them to those present here, since they could,
at first sight, be considered different, nevertheless they are
equivalent.

The so--called Gross--Pitaevskii--Poisson system is taken as the
fundamental point for the description of the situation. This means
(\ref{equa20}), or its equivalent version in the form of the Madelung
formalism, plus the equation explaining the behavior of the
gravitational potential $\Phi(\vec{r})$, i.e.,
$\nabla^2\Phi(\vec{r})=4\pi G\rho(\vec{r})$. The absence of
analytical solutions implies that an approximation has to be
introduced, and this is done by means of the Gaussian Ansatz
\citep{Chavanis1}. Afterwards, with this assumption the energy of the
system is deduced as well as all its ensuing consequences, among
them the size of the halo, etc. This is, in very few words, the
logic and conditions found in the literature in this context.

We now proceed to explain the assumptions used in the present work.
The gravitational effects, as explained before, are taken into
account in the form of an effective three dimensional harmonic
oscillator whose frequency hinges upon the mass and number of
particles in the BEC, see (\ref{equa13}), but not on the scattering
length. This fact entails not only that we do not need the
Gross--Pitaevskii--Poisson system, just the Gross--Pitaevskii
equation, but also that the situation is, in its mathematical
analysis, reduced to the case of a terrestrial BEC trapped by the
usual means \citep{Stringaribook1}.

The second assumption concerns the introduction of the scattering
length as a fundamental element of the description of the system.
This is done taking (\ref{equa20}) and noting that

\begin{eqnarray}
	\vert\psi(\vec{r})\vert^2\psi(\vec{r})=\rho(\vec{r})\psi(\vec{r})/m.\label{equa23}
\end{eqnarray}

Then the Gaussian structure for the density is introduced and
expanded in terms of a Taylor series, keeping terms up to second
order in $r$. The term depending quadratically upon $r$ is added to
the term emerging from the gravitational part, i.e., the term in
(\ref{equa22}) whose coefficient is $\omega_{(0)}$. The resulting
expression is then

\begin{eqnarray}
	\tilde{\mu}\psi(\vec{r})= -\frac{\hbar^2}{2m}\nabla^2\psi(\vec{r})+\frac{m\omega^2}{2}r^2\psi(\vec{r}),\label{equa24}
\end{eqnarray}

\begin{eqnarray}
	\omega^2=\omega^2_{(0)}-2\frac{NU_{(0)}}{m\sqrt{\pi^3}l^5},\label{equa25}
\end{eqnarray}

\begin{eqnarray}
	\tilde{\mu}=\mu-\frac{NU_{(0)}}{\sqrt{\pi^3}l^3}.\label{equa26}
\end{eqnarray}

Here $\tilde{\mu}$ is the effective chemical potential the one
suffers a modification due to the presence of the pairwise
self--interaction present in the system. Expression (\ref{equa24})
is our equation of motion for the analysis of the halo. Clearly, it
is a three--dimensional harmonic oscillator in which the frequency
is now modified due to the presence of a non--vanishing scattering
length. If we take a closer look at the study of the size of a BEC
trapped by a harmonic oscillator and in which the interaction among
the particles is introduced we may find that a perturbational
approach entails that the system corresponds to a harmonic
oscillator in which the corresponding frequency is modified due to
the presence of the scattering length \citep{Baym}. In this sense the
approximation encoded in (\ref{equa24})--(\ref{equa25}) is
equivalent to the perturbational approach found in the literature,
at least in the sense that both of them are three--dimensional
harmonic oscillators.

Let us now explain some additional differences between the usual
treatment \citep{Bohmer} and our model. In the literature around the
mathematical analysis of a static gravitational bounded BEC the
introduction of the Thomas--Fermi approximation plus the condition
of a polytropic equation of state ends up with the so--called
Lane--Emden equation \citep{Bohmer} the one implies a density
distribution for the dark matter provided by the spherical Bessel
function of $0$ order., i.e., $\rho_{dm}\sim\sin{(kr)}/(kr)$.

The Thomas-Fermi approximation requires the fulfillment of a
condition involving the total number of particles ($N$), the
scattering length ($a$) plus the characteristic length ($l$) related
to the trapping potential, i.e., $(Na)/l>1$ \citep{Pethickbook}. In
the current works in this context the use of Thomas--Fermi condition
is assumed, before the corresponding values for $N$, $a$, and $l$
are known. As mentioned before we do not resort to this
approximation since our goal is the deduction of these parameters
and, in consequence, we have no information allowing us to introduce
it. In this sense our approach is a more general one. At this point
a word of warning is required. Indeed, the Thomas--Fermi
approximation neglects the kinetic energy when compared to the
interaction or oscillator energies. In other words, there is no
kinetic energy in this approach. In our model the situation is
translated into the case of an effective harmonic oscillator and, in
consequence, due to the structure of the corresponding Hamiltonian
the expectation values of the potential and kinetic energies are
equal \citep{Cohenbook1}. This last argument tells us that if we
consider the case of vanishing kinetic energy in our procedure, then
we also demand implicitly (due to the fact that they are equal)
vanishing potential energy and therefore the whole energy vanishes.
In other words, the comparison against the case deduced resorting to
Thomas--Fermi cannot be done.

\section{Microscopic and macroscopic variables}

We now proceed to analyze the relationship between two different
sets of parameters related to the halo, namely, microscopic and
macroscopic features. The idea is to consider the conditions related
to the mechanical equilibrium of the halo, its relation to the
tangential velocity of rotating galaxies, and, finally, how a beam
of light is deflected by dark matter. These three aforementioned
conditions have a macroscopic character and will be cast as
functions of the microscopic ones, i.e., mass and scattering length
of the dark matter particles plus the number of particles contained
in the halo. In this manner we will have posed three expressions for
the three microscopic variables in terms of characteristics which
have an observational possibility.

\subsection{Macroscopic parameters}

\subsubsection{Mechanical equilibrium}

The first point concerns the expression defining the condition of
mechanical equilibrium for the BEC. Indeed, gravity tends to
collapse the halo and this behavior faces a pressure which is a
consequence of Heisenberg's Uncertainty Principle and of the
movement of the particles of the halo. Mechanical equilibrium
emerges when the corresponding pressures of these two processes are
equal.

The energy of the system due to $N_{(0)}$ particles in the ground
state is given by

\begin{eqnarray}
	E_{(0)}=\frac{3}{2}\hbar\omega N_{(0)}.\label{equa27}
\end{eqnarray}

In this last expression the frequency corresponds to (\ref{equa25})
and it does not neglect the kinetic energy, i.e., our formalism does
not resort to the Thomas--Fermi approximation \citep{Pethickbook}.

Since our system is equivalent to a terrestrial BEC trapped by a
three--dimensional isotropic harmonic oscillator then we must
remember that even in the case of vanishing temperature the presence
of $a\not=0$ implies, unavoidably, that excited states have to be
also populated, a physical consequence of the presence of a finite
coherence length \citep{Stringaribook1}.

The literature in the realm of dark matter as a BEC usually neglects
the effects of the thermal cloud \citep{Chavanis1, Bohmer}. The quest
for a more realistic formalism has led us to introduce the depletion
term which is the variable containing the information of the number
of particles populating excited states, namely

\begin{eqnarray}
	N_{(e)}=\frac{8}{3}N\left(\frac{Na^3}{\pi V}\right)^{1/2}.\label{equa28}
\end{eqnarray}

This last result stems from the case in which the BEC is trapped by
walls, not by a harmonic oscillator. In order to fathom better the
reasons for its use for cases in which a harmonic oscillator is
present let us comment that there is, approximately, one excited
particle per volume $\xi^3$, being $\xi$ the so--called coherence
length \citep{Pethickbook}. The definition of coherence length under
the presence of walls reads: it is the distance over which the wave
function rises from zero at the wall to close to its bulk value.
Mathematically this is written as: $\frac{\hbar^2m}{\xi^2}=
nU_{(0)}$. Under the presence of a trapping potential with the
structure of a harmonic oscillator we notice that this condition
shall include the energy of each particle emerging from the
interaction with the trap, namely: $\frac{\hbar^2m}{\xi^2}= nU_{(0)}
+m\omega^2(\xi-R)^2/2$. Assuming that the energies of the
self--interaction and that from the trap are similar in their order
of magnitude, $nU_{(0)}\sim m\omega^2(\xi-R)^2/2$, then
$\frac{\hbar^2m}{\xi^2}= 2nU_{(0)}$, and hence in this new situation
the coherence length has the same order of magnitude that in the
case of walls as trapping potential, as a matter of fact the
difference is only a factor of $\sqrt{2}$. Therefore the depletion
terms is provided by

\begin{eqnarray}
	N_{(e)}=\frac{2^{7/2}}{3}N\left(\frac{Na^3}{\pi V}\right)^{1/2}.\label{equa29}
\end{eqnarray}

This last argument proves that in a very simplified scheme the
structure of the coherence length, for the case in which a
non--trivial trapping potential is present, has the same structure
as the homogeneous situation.

These excited particles provoke pressure which has to be included in
the calculation of the mechanical equilibrium. Clearly, $N_{(e)}$
particles in the thermal cloud induce a larger pressure than the
same number of particles in the ground state, the reason is related
to the fact that in excited states they have a larger momentum.
Therefore we may conjecture that the inclusion of the depletion term
will imply, among other possibilities, halos with larger masses than
those appearing in works in which all particles populate the ground
state. In addition, the halo will also have a larger volume, here
the argument lies upon the fact that excited states related to bound
particles occupy a larger region than those in the ground state. In
other words, the presence of the depletion term allows us to
conjecture that here we will obtain larger and more massive halos.

In order to obtain the energy of those particles in excited states
we resort to (\ref{equa24}) which describes the BEC as a Hamiltonian
with an effective three-dimensional harmonic oscillator potential.
Of course, a particle in the first excited state of this system will
have the energy $\epsilon_{(1)}=\frac{5}{2}\hbar\omega$ and, in
consequence, our approximation is that the thermal cloud has an
energy equal to

\begin{eqnarray}
	E_{(1)}=\frac{5}{2}\hbar\omega N_{(e)}.\label{equa30}
\end{eqnarray}

The total energy is provided by

\begin{eqnarray}
	E_{(T)}= E_{(0)} + E_{(1)}.\label{equa31}
\end{eqnarray}

The pressure due to this energy is given by the expression $P_{(c)}
= -\frac{\partial E_{(T)}}{\partial V}$ and since we have a
spherically symmetric body of volume $V$ this parameter is a
function of the characteristic length of our harmonic oscillator and
we must deduce it. In order to do this we consider the content of
dark matter at those points at which the rotational speed of a
galaxy takes its maximum value, namely, the size of a sphere such
that any star located on its surface has a rotational speed whose
derivative (with respect to the distance to the geometric center of
the halo) vanishes. The physical reason for this particular choice
will be thoroughly explained in the next subsection in connection
with one of our extant astronomical readouts.

The functional relation between speed and dark matter is given by

\begin{eqnarray}
	v^2(r)=\frac{GM(r)}{r}.\label{equa32}
\end{eqnarray}

Here $M(r)$ denotes the total dark matter contained in a sphere of
radius $r$, namely, $M(r) = M_{(0)}(r)+M_{(e)}(r)$, the sum of the
mass of particles in the ground state and excited states,
respectively. Since our model is, mathematically, a
three--dimensional harmonic oscillator, then (assuming that only the
first excited state is populated) we have that

\begin{eqnarray}
	M_{(0)}(r)=\frac{4mN_{(0)}}{\sqrt{\pi}l^3}\int_0^rz^2\exp{(-z^2/l^2)}dz,\label{equa33}
\end{eqnarray}

\begin{eqnarray}
	M_{(e)}(r)=\frac{8mN_{(e)}}{3\sqrt{\pi}l^5}\int_0^rz^4\exp({-z^2/l^2)}dz.\label{equa34}
\end{eqnarray}

The condition $dv^2/dr=0$ becomes

\begin{eqnarray}
	M_{(0)}(r=R)+M_{(e)}(r=R)=\frac{4m}{\sqrt{\pi}l^3}\Bigl(N_{(0)}R^3\nonumber\\
	+\frac{2N_{(e)}}{3l^2}R^5\Bigr)\exp{(-R^2/l^2)}.\label{equa35}
\end{eqnarray}

Integration by parts of expressions (\ref{equa33}) and
(\ref{equa34}) entails that (\ref{equa35}) is equivalent to

\begin{eqnarray}
	\left(N_{(0)}(R/l)^3+\frac{2}{3}N_{(e)}(R/l)^5\right)\exp{(-R^2/l^2)}=\nonumber\\
	-\left(\frac{N}{2}(R/l)+\frac{N_{(e)}}{3}(R/l)^3\right)\exp{(-R^2/l^2)}\nonumber\\
	+\frac{N}{2l}\int_0^R\exp({-z^2/l^2)}dz.\label{equa36}
\end{eqnarray}

Concerning the integral in this last expression we may comment that
it is related to the so--called probability integral ($\Phi(z)$) and
among its possible representations we have the following one
\citep{Grad}

\begin{eqnarray}
	\Phi(z)=\frac{2}{\sqrt{\pi}}\exp(-z^2)\sum_{s=0}^{\infty}\frac{2^sz^{2s+1}}{(2s+1)!!}.\label{equa37}
\end{eqnarray}

In order to have an analytical expression we consider in the last
 series up to $s=3$. The theory of ultra--cold dilute bosonic gases tells us that
$N_{(0)}>N_{(e)}$, introducing this condition in our ensuing
algebraic equation implies that the equation to be solved reads

\begin{eqnarray}
	-\left(\frac{R}{l}\right)^3+\frac{1}{5}\left(\frac{R}{l}\right)^5+\frac{2}{35}\left(\frac{R}{l}\right)^7=0. \label{equa38}
\end{eqnarray}

The solution is
\begin{eqnarray}
	R=1.67l.\label{equa39}
\end{eqnarray}

Therefore the content of dark matter within the sphere of radius $R$
is

\begin{eqnarray}
	M(r=R)=\frac{4mN_{(0)}}{\sqrt{\pi}l^3}\Bigl(\frac{l^3}{2}\int_0^{1.67}\exp(-w^2)dw\nonumber\\
	-\frac{(1.67)}{2}(0.061)l^3\Bigr)\nonumber\\
	+\frac{8mN_{(e)}}{3\sqrt{\pi}l^5}\Bigl(\frac{3l^5}{4}\int_0^{1.67}\exp(-w^2)dw\nonumber\\
	-\frac{(14.3)}{4}(0.061)l^5\Bigr).\label{equa40}
\end{eqnarray}

Resorting to tables \citep{Abra} we obtain that

\begin{eqnarray}
	\int_0^{1.67}\exp(-w^2)dw=0.87.\label{equa41}
\end{eqnarray}

Finally, the sought mass reads

\begin{eqnarray}
	M(r=R)=(0.87)mN\left[1-\frac{2^{3/2}}{3}\sqrt{\frac{Na^3}{R^3}}\right].\label{equa42}
\end{eqnarray}

We now hark back to the deduction of the pressure of the condensate
such that the volume is define as $V=\frac{4\pi}{3}(1.67l)^3$. The
ensuing result is

\begin{eqnarray}
	P_{(c)}= \frac{6\hbar^2N}{13.4\pi mR^5}+\frac{6U_{(0)}N^2}{13.4\pi^{5/2}R^6}\nonumber\\
	+\frac{14\hbar^2N}{10\pi mR^5}\sqrt{\frac{2Na^3}{\pi V}}.\label{equa43}
\end{eqnarray}

It has to be underlined that the deduction of the energy of the
first excited state as a consequence of a perturbation process
related to our three--dimensional harmonic oscillator requires an
explanation. Clearly, the effects of the presence of a
non--vanishing scattering length for the ground state can be handled
according to a perturbation approach as long as the interaction
energy per particle in the ground state is smaller than the ground
energy associated to the unperturbed three--dimensional harmonic
oscillator Hamiltonian i.e.,

\begin{eqnarray}
	\vert<0,0\vert\hat{W}\vert0,0>\vert< \frac{3\hbar\omega}{2}.\label{equa44}
\end{eqnarray}

In this last expression $\hat{W}$ denotes the operator related to
the interaction potential of two particles in the ground state.
Concerning the first excited state we have the same kind of
condition behind this approximation

\begin{eqnarray}
	\vert<1,1\vert\hat{W}\vert1,1>\vert< \frac{5\hbar\omega}{2}.\label{equa45}
\end{eqnarray}

Let us now address the issue concerning the gravitational attraction
of the halo, a topic related to the equilibrium of a spherical body
with density, pressure, velocity field, and gravitational potential
$\rho$, $P$, and $\vec{v}$, $\Phi$, respectively. The corresponding
equations for the internal structure are \citep{Willbook}

\begin{eqnarray}
	\rho\frac{d\vec{v}}{dt}= \rho\nabla\Phi-\nabla P,\label{equa46}
\end{eqnarray}

\begin{eqnarray}
	\frac{\partial\rho}{\partial t}+\nabla\cdot(\rho\vec{v})=0.\label{equa47}
\end{eqnarray}

To close the system an additional piece of information is required,
namely, the equation of state, i.e., the functional dependence among
pressure, density, temperature, etc. The Newtonian gravitational
potential for a spherical body of radius $R$ is

\begin{eqnarray}
	\nabla\Phi(t,r)=\frac{GM(t,r)}{r}+4\pi\int_r^R\rho(t,r')dr'.\label{equa48}
\end{eqnarray}

In this last expression $M(t,r)$ is the mass contained in a sphere
(coincident with our body) of radius $r$. We now consider the
surface of our sphere and calculate the change in this gravitational
potential due to a change in the volume, a process that entails a
pressure ($P_{(g)}=-\frac{\partial\Phi}{\partial V}$)

\begin{eqnarray}
	P_{(g)}=\frac{GM^2}{4\pi R^4}.\label{equa49}
\end{eqnarray}

In the general situation the pressure is a non--constant function of
the radial distance \citep{Willbook} and, in consequence, we must
identify the value of $r$ related to (\ref{equa49}). Since this
pressure is deduced after evaluating the gravitational potential on
the surface of the body then it represents the pressure on this
surface. The mathematical condition determining mechanical
equilibrium is the equality of our two pressures on the surface of
the halo (remember that $M=mN$) i.e., expressions (\ref{equa43}) and
(\ref{equa49}).

One consequence of the equality $P_{(g)}=P_{(c)}$ is related to the
fact that we have deduced $R$ as a function of $m$, $a$, and $N$. In
other words, if we have the size of a halo, by mean of any kind of
astrophysical or astronomical observations, then we have the first
of our required expressions. These last comments also provide a
physical explanation to the choice done in the deduction of our
parameter $R$. Indeed, we are forced, due to the extant astronomical
observations, to deduce the content of matter and size (as function
of $l$) of the region at which the derivative of $v^2(r)$ vanishes.

\subsubsection{Concordance with rotation velocities}

It can be said, without exaggerating, that dark matter is a
descendant of the rotation curves of spiral galaxies \citep{Binney}
and, in consequence, we must include the explanation of them, as a
fundamental part of the present work. In this section we deduce the
relationship between these curves and the dark matter content of our
model.

Let us consider the tangential velocity of a star, which according
to Newtonian physics is provided by

\begin{eqnarray}
	\frac{v_t^2(r)}{r}= \frac{GM^{(T)}(r)}{r^2}.\label{equa50}
\end{eqnarray}

In this last expression $M^{(T)}(r)$ denotes the total mass within a
sphere of radius $r$, i.e., it includes baryonic and dark matter.
Let us now consider the density of the involved baryonic mass of the
halo ($\rho^{(b)}$) and dark matter in its ground state and excited
states, $\rho^{(0)}$, $\rho^{(e)}$, respectively (here $\rho^{(e)}$
is deduced from the wave function related to the first excited state
of a three--dimensional harmonic oscillator \citep{Cohenbook1})

\begin{eqnarray}
	M^{(T)}(r)=M^{(b)}(r)+\nonumber\\
	\int_0^r\int_0^{\pi}\int_0^{2\pi}\left[\rho^{(0)}(r')+\rho^{(e)}(r')\right]r'^2dr'd\Omega ,\label{equa51}
\end{eqnarray}

\begin{eqnarray}
	M^{(b)}(r)=\int_0^r\int_0^{\pi}\int_0^{2\pi}\rho^{(b)}(r')r'^2dr'd\Omega.\label{equa52}
\end{eqnarray}

Expression (\ref{equa50}) implies that

\begin{eqnarray}
	\frac{d\bigl(v_t^2(r)\bigr)}{dr}=-\frac{v_t^2(r)}{r}+\frac{G}{r}\frac{d\bigl(M^{(T)}(r)\bigr)}{dr}.\label{equa53}
\end{eqnarray}

Let us now denote by $r=R$ (here $R=1.67l$, see expression
(\ref{equa39})) the value of the radial coordinate at which
(\ref{equa53}) vanishes. Then we obtain the following condition

\begin{eqnarray}
	v_t^2(R)= 4\pi\rho^{(b)}(R)R^2 +\frac{4\pi mNG}{(\sqrt{\pi}l)^3}\biggl[1\nonumber\\
	+\sqrt{\frac{32a^3N}{3\pi^2l^3}}\left(-1+\frac{2R^2}{3l^2}\right)\biggr]R^2\exp{\left(-\frac{R^2}{l^2}\right)}.\label{equa54}
\end{eqnarray}

A fleeting glimpse at this last expression allows us to understand
that if we know the value of the radial coordinate at which the
tangential velocity has its maximum value, then have a second
expression (which is detectable) and hinges upon $m$, $a$, and $N$.
In other words, up to now we have two parameters which can be
measured and such that they are functions of the variables of the
BEC. The last required expression is the deflection of a light beam
due to the content of dark matter.

\subsubsection{Light deflection and BEC}

The goal here is to determine the deflection of a light beam due to
the presence of dark matter. For a spherical body with total mass
$M^{T}$, in the weak--field limit, the line element for points
outside the body reads \citep{Will93}

\begin{eqnarray}
	ds^2=-\left(1-2\frac{M^{(T)}}{r}\right)dt^2\nonumber\\
+\left(1+2\frac{M^{(T)}}{r}\right)\left(dx^2+dy^2+dz^2\right).\label{equa55}
\end{eqnarray}

The origin of the coordinate system coincides with the geometrical
center of our spherically symmetric halo. For the sake of clarity we
now consider a light beam moving in the plane $z=0$ and such that it
approaches the halo coming from $x\rightarrow-\infty$. The initial
conditions entail that for the momentum of the beam we have
$p^{(z)}=0$, $\vert p{(y)}\vert <<\vert p^{(x)}\vert$, and
$p^{(0)}\approx p^{(x)}$, along the whole trajectory. The impact
parameter is $y=R$, in other words, the beam, concerning its
position on the $y$--axis has a distance equal to the size of the
halo. Under these conditions the deflection angle reads
\citep{MTWbook}

\begin{eqnarray}
	\Delta\phi=-\frac{4GM^{(T)}}{c^2R}.\label{equa56}
\end{eqnarray}

In this last expression $M^{(T)}$ denotes the total mass, i.e., it
includes the baryonic contribution plus dark matter.

The contribution to the deflection angle stemming solely from dark
matter ($\Delta\phi^{(dm)}$) is easily calculated noting that the
rotation velocity at $r=R$ satisfies the relation $v^2(r=R)
=GM^{(dm)}/R$ \citep{Blanford}. Since the mass content within the
sphere of radius $R$ has been already calculated, see
(\ref{equa42}),  then the corresponding deflection angle entails

\begin{eqnarray}
	\frac{3.48mNG}{c^2R}= 2\pi\frac{v^2(r=R)}{c^2}.\label{equa57}
\end{eqnarray}

Clearly, this is a third expression relating $m$, $N$, and $a$ with
a parameter that can be detected astronomically, the right--hand
side of our last result.

\section{Discussion of results and conclusions}

We now resort to our three main expressions, i.e., the equality
between (\ref{equa43}) and (\ref{equa49}), plus (\ref{equa54}) and
(\ref{equa57}), which are functions of $m$, $a$, and $N$. Clearly,
we may deduce the aforementioned physical variables associated to
the BEC from our work. Indeed, we have three unknowns and three
equations.

In order to obtain these parameters we will use the values of the
rotation velocities of some of the stars reported in the SPARC (Spitzer Photometry \& Accurate Rotation Curve) data set \citep{Lelli}.
The corresponding values are given in Table \ref{tbl.1}.

\begin{table*}
	\caption{\label{tbl.1}}
	\begin{tabular}{@{}lccccccc@{}}
		\tableline
		\tableline
		Galaxy & $v$ & $R$ & $Nm$ & $m$ & $N$ & $a$ & $(V/N)^{1/3}$ \\
		&  &  & $10^{9}$ &  & $10^{73}$ & $10^{-6}$ & $10^{-5}$ \\
		& [km/s] & [kpc] & [$\mathrm{M}_{\odot}$] & [$\mathrm{eV}/c^{2}$] & & [m] & [m] \\
		\tableline
			N0055 & 86.8 & 9.82 & 31.11 & 21.44 & 161.6 & 7.61 & 4.16 \\
			N0300 & 97.0 & 9.53 & 37.70 & 21.17 & 198.3 & 6.90 & 3.77 \\
			N1090 & 176.0 & 11.28 & 146.90 & 16.77 & 975.9 & 4.80 & 2.63 \\
			N3198 & 157.0 & 14.05 & 145.60 & 15.46 & 1049.1 & 5.84 & 3.19 \\
			N6015 & 166.0 & 21.04 & 243.76 & 12.46 & 2179.4 & 6.85 & 3.75 \\
			U8550 & 57.8 & 4.39 & 6.17 & 35.50 & 19.3 & 6.91 & 3.78 \\
			U9037 & 160.0 & 19.46 & 209.45 & 13.07 & 1784.5 & 6.78 & 3.70 \\
			DDO64 & 46.4 & 2.08 & 1.88 & 54.49 & 3.9 & 5.61 & 3.06 \\
			U5005 & 97.1 & 17.97 & 71.23 & 15.41 & 514.8 & 9.47 & 5.18 \\
			U5750 & 77.6 & 11.43 & 28.94 & 20.44 & 157.7 & 8.93 & 4.88 \\
			U731 & 73.4 & 8.19 & 18.55 & 24.49 & 84.4 & 7.89 & 4.31 \\
			N2366 & 53.7 & 4.16 & 5.04 & 37.15 & 15.1 & 7.10 & 3.88 \\
			N3274 & 82.6 & 1.89 & 5.42 & 49.49 & 12.2 & 3.47 & 1.90 \\
			N5023 & 82.9 & 6.15 & 17.77 & 27.41 & 72.2 & 6.24 & 3.41 \\
		\tableline
	\end{tabular}
\end{table*}

In addition, a possible situation is related to the case in which
there are more than one different kind of dark matter particles, a
possibility that cannot be discarded from the very beginning.
Nevertheless, if we assume that all these fourteen galaxies contain dark
matter stemming from the same kind of particle, then the deduced
values (for $a$ and $m$) shall not have a large spreading, in a
statistical sense.

In order to have a deeper comprehension of the present predictions
we proceed to calculate the standard deviation ($\sigma$) for $m$
and $a$ as well as the ratio between the standard deviation and
square of the average value of the corresponding parameter (see Table \ref{tbl.2}).

\begin{table*}
	\caption{\label{tbl.2}}
	\begin{tabular}{@{}lcccc@{}}
		\tableline
		\tableline	
		 & Mean $\overline{x}$ & Standard Deviation $\sigma$ & $(\sigma/\overline{x})$ & $(\sigma/\overline{x})^2$\\
		\tableline
		$m$ & $26.05$ $\mathrm{eV}/\mathrm{c}^2$ & $12.86$ & $0.50$ & $0.24$\\
		$a$ & $6.74$ x $10^{-6}$m & $1.49$ x $10^{-6}$ & $0.22$ & $0.05$\\
		\tableline
	\end{tabular}
\end{table*}

It is readily seen that the statistical error,
corresponds to a twenty four percent for the mass of the dark matter
particle, whereas for the scattering length it is almost five times
smaller, i.e., five percent. In other words, the model provides
consistent values for $m$ and $a$ in the sense that the predictions
for fourteen different galaxies do not show a large spreading.

We underline the fact that in the present work we do not introduce
by hand a value for any of the microscopic properties of the BEC,
they are rather calculated.

A second test of the validity of our approach can be obtain noting
that the formalism codified in our fundamental expression
(\ref{equa19}) requires the fulfillment of the condition of dilute
gas, otherwise interactions involving more than two particles would
become relevant in the description of the system. This condition is
given by

\begin{eqnarray}
	\left(\frac{V}{N}\right)^{1/3}>a.\label{equa58}
\end{eqnarray}

Our calculations show that the condition is satisfied, as can be
seen from our tables. There is another inequality to be fulfilled.
Indeed, we know that $N\geq N_{(e)}$, resorting to according to
(\ref{equa29}) we have that
$(3\sqrt{\pi}/2^{7/2})N_{(e)}=N(Na^3/V)^{1/2}$ and this condition
implies that $(3\sqrt{\pi}/2^{7/2})\geq (Na^3/V)^{1/2}$. A fleeting
glimpse to the results here contained proves that our values do
fulfill this aforementioned condition.

As mentioned before the Thomas-Fermi approximation requires the
fulfillment of the condition $(Na)/R>1$ \citep{Pethickbook}, the one
involves not only $N$ but also two additional parameters. Stating
that a large number of particles is tantamount to Thomas--Fermi is a
mistake. In the literature of the relationship dark matter--BEC the
validity of Thomas--Fermi condition is assumed, before the
corresponding values for $N$, $a$, and $R$ are known. As mentioned
before we do not resort to this approximation since our goal is the
deduction of these parameters and, in consequence, we have no
information allowing us to introduce it.

Summing up, we modeled dark matter as a Bose--Einstein condensate
and considered the effects of the thermal cloud. The effects of the
self--gravitational interaction have been reformulated in terms of a
three--dimensional harmonic oscillator and the pairwise interaction
present in the case of small energies has also been considered,
without resorting to the Thomas--Fermi approximation. The main goal
has been the deduction of the microscopic properties, namely, mass,
number of particles, and scattering length, related to the particles
comprised in the corresponding condensate. This task has been
achieved introducing three macroscopical physical conditions related
to the halo, i.e., mechanical equilibrium of the condensate,
explanation of the rotation curves of stars belonging to LSB
galaxies, and, finally, the deflection of light due to the presence
of dark matter. These three aforementioned expressions allowed us to
cast the features of the particles in terms of detectable
astrophysical variables. Finally, the model has been contrasted
against the observational data of fourteen galaxies and in this manner
we obtained values for the involved microscopic parameters of the
condensate. The deduced values show an error of twenty four percent for
the mass of the dark matter particle and five percent in connection
with the scattering length.

We may wrap up our discussion stating that the present proposal
provides a deduction, within the realm of the theory of dilute
ultra--cold gases, for the microscopic properties of a dark matter
halo. Additional features of physical relevance remain to be
analyzed in this context and the corresponding results will be
published elsewhere.


%
%

%
%

%

%
%

%

%
\acknowledgments
B. C. acknowledges CONACyT grant No. 574365 and S. G. the received
UAM grant.


%

%

\end{document}